\documentclass[epj]{svjour}
\usepackage{graphics}
\begin{document}
\title{Phase transition of surface models with intrinsic curvature}
\author{H. Koibuchi \and N. Kusano \and A. Nidaira \and Z. Sasaki \and K. Suzuki% etc
%\thanks{\emph{Present address:Nakane 866, Hitachinaka, Ibaraki 312-8508, Japan } }%
}                     % Do not remove
%
%\offprints{koibuchi@mech.ibaraki-ct.ac.jp}          % Insert a name or remove this line
%
\institute{Department of  Mechanical and Systems Engineering, \\
  Ibaraki College of Technology \\
  Nakane 866, Hitachinaka, Ibaraki 312-8508, Japan }
%
%\date{Received:  Revised version:  }
\abstract{
It is reported that a surface model of Polyakov strings undergoes a first-order phase transition between smooth and crumpled (or branched polymer) phases. The Hamiltonian of the model contains the Gaussian term and a deficit angle term corresponding to the weight of the integration measure $dX$ in the partition function.}
 %end of abstract
\maketitle
%

%--------------------------------------------
\section{Introduction}
%--------------------------------------------
A large effort has been devoted to the understanding of the phase structure of elastic membrane models of Polyakov-Kleinert \cite{POLYAKOV,Kleinert} and Helfrich \cite{HELFRICH}. The surface models are expected to undergo phase transitions between smooth and crumpled phases at finite bending rigidity \cite{DavidGuitter,Peliti-Leibler,BKS,BK,Kleinert-2,David,NELSON,KANTOR,BOWICK-TRAVESSET,WIESE,WHEATER-1}. Numerical studies have also been concentrated on the phase transition \cite{WHEATER-2,BOWICK-1,AMBJORN,CATTERALL,BOWICK-2,ANAGNOST,KANTOR-NELSON,GOMPPER-KROLL,KOIB-PLA-2002,KOIB-PLA-2003-2}.

The bending energy, which is an extrinsic curvature term, plays an essential role in smoothing the surface.  In the ordinary model of membranes, the smooth phase emerges owing to the bending energy. Except self-avoiding surface models \cite{ARP,Ho-Baum,Abraham-Nelson,Grest,BOWICK-3}, we have currently no surface model that has a smooth phase without the extrinsic curvature term. 

 Baillie-Johnston et. al. \cite{Baillie-Johnston,BEJ,BIJJ,Ferguson-Wheater} investigated the discretized Polyakov random surface model with intrinsic curvature terms in order to find smooth surfaces. One of the models was recently investigated in \cite{KOIB-PRE-2003} on relatively larger lattices, however, there was no smooth phase in the model, although branched polymer like phase was found. We consider that the question about a smooth phase in the models remains unanswered, and the surface model with intrinsic curvatures remains to be studied. Thus, motivated by the attempts of Baillie-Jhonston et. al., we study in this article a model with an intrinsic curvature term and see that the smooth phase and the crumpled phase are appeared and connected by a phase transition. Our results may lead to better understanding of the role of curvature energy in surface models undergoing phase transitions and/or shape transformation, although the surfaces are allowed to self-intersect and hence phantom.

Two kinds of models are investigated; the tethered model and the fluid model. The tethered model is defined on fixed connectivity surfaces, and the fluid model on dynamically triangulated surfaces. The Hamiltonians of the models are identical with each other. It will be shown that the tethered model has the smooth and the crumpled phases, which are connected by a first-order phase transition. Moreover, we will find in the fluid model that there appears a new phase, which resembles the branched polymer phase \cite{Gom-Krol-2,Boal-Rao}, between the smooth phase and the crumpled one. We understand from the results obtained in this article that the smooth phase is stable against such as thermal fluctuations in each model defined only with intrinsic variables of the surface. 

%--------------------------------------------
\section{Model}
%--------------------------------------------
The partition function of the fluid model of fixed number of vertices is defined by
\begin{eqnarray} 
\label{Part-Func}
 Z = \sum_{\cal T} \int \prod _{i=1}^{N} d X_i \exp\left[-S(X,{\cal T})\right],\\  
 S(X,{\cal T})=S_1 - \alpha S_3, \nonumber
\end{eqnarray} 
where the center of the surface is fixed to remove the translational zero mode. $S(X,{\cal T})$ denotes that the Hamiltonian $S$ depends on the position variables $X$ of vertices and the degree of freedom for the triangulation ${\cal T}$. $S_1$ and $S_3$ are defined by
\begin{equation}
\label{Disc-Eneg} 
S_1=\sum_{(ij)} \left(X_i-X_j\right)^2,\quad S_3=\sum_i\log( \delta_i/2\pi),
\end{equation} 
where $\sum_{(ij)}$ in $S_1$ is the sum over bonds $(ij)$ connecting the vertices $i$ and $j$. The bonds $(ij)$ are edges of the triangles. $\delta_i$ in Eq. (\ref{Disc-Eneg}) is the vertex angle, which is the sum of the angles meeting at the vertex $i$. Recalling that $\delta_i\!-\!2\pi$ is the deficit angle, we call $S_3$ the deficit angle term. We note that the sum of the deficit angles $\sum_i (\delta_i\!-\!2\pi)$ is constant on surfaces of fixed genus because of the Gauss-Bonnet theorem. The partition function of the tethered model is, on the other hand, defined by $Z \!=\!  \int \prod _{i=1}^{N} d X_i \exp\left[-S(X,{\cal T})\right]$, where ${\cal T}$ is fixed. Both $S_1$ and $S_3$ reflect intrinsic properties of the surface. Note also that $S_1/N=3/2$, which comes from the scale invariance property of $Z$ in both models.

We should refer to a relation between the deficit angle term $S_3$ and the integration measure $dX_i$ of the partition function in Eq. (\ref{Part-Func}). From a conformal field theoretical viewpoint \cite{David-NP,BKKM,ADF,FN}, the integration measure $\prod_i dX_i$ can be replaced by  $\prod_i q_i^{\alpha} dX_i$, where $q_i$ is the co-ordination number of the vertex $i$. This $\alpha$ is believed to be $2\alpha\!=\!3$. On the other hand, $q_i^{\alpha}$ is considered as a volume weight of the vertex $i$ in the measure $dX_i$. Hence, it is possible to extend $2\alpha$ to non-integer numbers by assuming that the weight can be chosen arbitrarily. Moreover, it is also possible to extend $q_i$ to continuous numbers because of the same reason. Hence, the weight $\prod_i q_i^{\alpha}$  can be replaced by $\prod_i \delta_i^{\alpha}$, which can also be written as $\exp ( \alpha \sum_i \log \delta_i )$. Including a constant weight $(2\pi)^\alpha$, we have $S_3$ in Eq. (\ref{Disc-Eneg}). 

The model in this paper is very similar to the one investigated in \cite{Baillie-Johnston,BEJ,BIJJ,Ferguson-Wheater,KOIB-PRE-2003}, where the fluid and the tethered models were interpolated. A model in \cite{Baillie-Johnston,BEJ,BIJJ,Ferguson-Wheater,KOIB-PRE-2003} includes the term $\alpha \sum_i |q_i-6|$ or $\alpha \sum_i (q_i-6)^2$. Using the expression $\log (\delta_i/2\pi) \!=\! \log [1 \!+\!(\delta_i\!-\!2\pi)/2\pi]$ and replacing $\delta_i$ by $q_i$ and $2\pi$ by 6  in  $S_3$ of Eq. (\ref{Disc-Eneg}), we have the term proportional to  $\sum_i (q_i-6)^2 + ({\rm higher \; order})$. Thus, we confirmed that the lowest order term in $S_3$ is identical with the expression $\alpha \sum_i (q_i-6)^2$ described above upto multiplicative constant. We note that the similarity between  $S_3$ of Eq. (\ref{Disc-Eneg}) and a deficit angle term denoted by $S_{\rm tight}$ in \cite{Baillie-Johnston} can also be seen.  

 We will see below that the fluid surfaces become crumpled, branched polymer like, and smooth, when  $\alpha$ increases from $\alpha\!=\!0$. A higher-order transition separates the branched polymer phase from the crumpled one, and a first-order transition separates the branched polymer phase from the smooth one. On the other hand, we find in \cite{KOIB-PRE-2003} that the fluid surfaces become crumpled, branched polymer, and crumpled, with increasing $\alpha$. One of the transitions is of  higher (or second) order, and the other is of first order, in the model of \cite{KOIB-PRE-2003}. 

It should also be noted on differences between $S_3(\delta)$ of Eq. (\ref{Disc-Eneg}) and the corresponding term $S_3(q)\!=\!\sum_i\log (q_i/6)$, which can be expanded to $\sum_i (q_i-6)^2 + ({\rm higher \; order})$ as described above. Although the difference between these two $S_3$ comes only from the one between $q$ and $\delta$, the roles of $q$ and $\delta$ in the model make large differences in the results. A major difference between the fluid model in this paper and the one in \cite{KOIB-PRE-2003} is whether the model has a smooth phase or not. $q$ is unchanged  in the MC update of $X$, whereas $\delta$ changes in that process.  As a consequence, while the co-ordination dependent term $S_3(q)$ is non-trivial only in the fluid model, $S_3(\delta)$ in Eq. (\ref{Disc-Eneg}) is non-trivial both in the tethered and in the fluid models. In fact, the model with $\sum_i (q_i-6)^2$ has no smooth phase \cite{KOIB-PRE-2003}, whereas the model with $S_3(\delta)$ in Eq. (\ref{Disc-Eneg}) has the smooth phase both in the tethered and in the fluid surfaces as we will see below.   

%--------------------------------------------
\section{Monte Carlo Simulations}\label{MC-simul}
%--------------------------------------------
The spherical surfaces, the starting configurations of MC for the fluid model, are discretized by the Voronoi triangulation \cite{FRIEDBERG-REN} in ${\bf R}^3$. The radii of the initial spheres are chosen so that $S_1/N\!\simeq\!3/2$. The surfaces are constructed relatively uniform in co-ordination number; the maximum co-ordination number is about 8.  

On the contrary, the uniform lattices, on which the tethered model is defined, were obtained by MC with dynamical triangulation for a model defined by $S\!=\!S_1\!+\!bS_2\!-\!\alpha \sum_i \log ( q_i/6)$, where $S_2\!=\!\sum_i(1\!-\!\cos\theta_i)$ is the bending energy. Using the parameters $\alpha\!=\!280$, $b\!=\!2$ (or $\alpha\!=\!350$, $b\!=\!2$), we obtained the uniform lattices. By monitoring the number $N_5$ of vertices of $q_i\!=\!5$, we stopped the MC run when $N_5$ first reduced to 12, and thus obtained the uniform lattices. The uniform lattices, which will be utilized in simulations for the tethered model, are thus characterized by $N_5\!=\!12$ and $N_6\!=\!3N\!-\!18$. In each lattice, only 12 vertices are of $q_i\!=\!5$, and all other vertices are of $q_i\!=\!6$. Except the lattices of $N_5\!=\!12$, there exists no lattice whose $q_i$ are of $q_i\!=\!5$ or $q_i\!=\!6$. This is easily understood from the fact that $N_{\rm B}\!=\!3N\!-\!6$ on triangulated surfaces of spherical topology, where $N_{\rm B}$ is the total number of bonds. 

The canonical Metropolis MC technique is used to update the variables $X$ and ${\cal T}$. In the MC simulations, the new position of $X$ is chosen so that $X^\prime \!=\! X \!+\! \delta X$, where $\delta X$ takes value randomly in a small sphere. The radius $\delta r$ of the small sphere is defined by using a constant number $\epsilon$ as an input parameter so that
\begin{equation}
\label{epsilon}
 \delta r = \epsilon\, \langle l\rangle,
\end{equation}
 where $\langle l\rangle$ is the mean bond length computed at every 250 MCS (Monte Carlo sweeps), and $\langle l\rangle$ is constant in the equilibrium configurations because $S_1/N\!=\!3/2$. The new position $X^\prime $ is accepted with the probability \\ ${\rm Min} [1,\exp(-{\mit \Delta} S)]$, where ${\mit \Delta} S\!=\!S({\rm new})\!-\!S({\rm old})$. The parameter $\epsilon $ in Eq. (\ref{epsilon}) is chosen to maintain about $50\%\!\sim\!60\%$ acceptance for $X$. A lower bound $10^{-6}\!\times\!A_0$ is assumed for the area of triangles, where $A_0$ is the mean area of triangles evaluated at every 250 MCS. $X^\prime$ is generated so that the resulting areas of triangles are larger than $10^{-6}\!\times\!A_0$. It should be emphasized that the results obtained by using $\delta r$ in Eq. (\ref{epsilon}) are identical with those by using a constant $\delta r$.

 The triangulation ${\cal T}$ is updated by the bond flip. Then, the acceptance rate $r_{\cal T}$ for the bond flip is uncontrollable. We have, in our MC simulations, $15\% \leq r_{\cal T} \leq 35 \%$, and find that $r_{\cal T}$ discontinuously changes at a discontinuous transition which will be shown below.

  The bond is flipped so that the resulting areas of triangles are larger than $10^{-6}\!\times\!A_0$ as in the process for $X$. No restriction is imposed on the bond length in the update of the variables $X$ and ${\cal T}$. The maximum co-ordination number $q_i^{\rm max}$ of the fluid surface of size $N\!=\!1500$ was about $q_i^{\rm max}\!\simeq\!20$ both in the smooth phase and in the branched polymer phase, and $q_i^{\rm max}\!\simeq\!30$ in the crumpled phase. The histogram $h(q)$ of the co-ordination number $q$ will be shown below. $N$-trials for $X$ and $N$-trials for ${\cal T}$ make one MCS in the fluid model. 

 The canonical ensemble average $\langle Q\rangle$ is computed via the time series average of a series $\lbrace Q_i\rbrace$.  A series  $\lbrace Q_i\rbrace$ is sampled at every 500 MCS after thermalization sweeps, which are large enough so that the mean square size $X^2$, defined by
\begin{equation} 
X^2=\frac{1}{N} \sum_i \left( X_i -\bar X \right)^2,\qquad \bar X =\frac{1}{N} \sum_i X_i,
\end{equation} 
 becomes statistically independent from the starting configurations. In fact, $0.5\times10^7$ MCS are discarded for the thermalization of surfaces of $N\!\leq\!600$, and $1\times10^7$ MCS for $N\!\geq\!1000$ surfaces. Total number of MCS is about $(8\sim 10)\times 10^7$ at the vicinity of the transition point of $N\!\geq \!1000$ fluid model and of  $N\!\geq \!1500$  tethered model, and it is relatively smaller in other cases.

%%%%%%%%%%%%%%%%%%%%%%%%%%%%%%%%%%%%%%%%%%%%%%%%%%%%%%%%%%%
\begin{figure}[htb]
\resizebox{0.49\textwidth}{!}{%
  \includegraphics{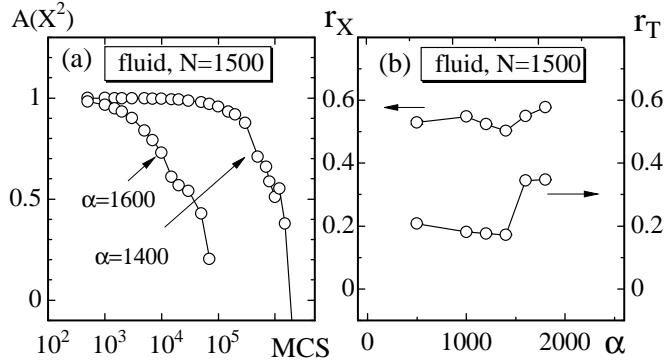}
}
 \caption{(a) $A(X^2)$ at $\alpha\!=\!1400$ and at $\alpha\!=\!1600$ on $N\!=\!1500$ fluid surface, and (b) the acceptance rates $r_X$ for $X$ and $r_T$ for ${\cal T}$ of the fluid surface model.} 
\label{fig-1}
\end{figure}
%%%%%%%%%%%%%%%%%%%%%%%%%%%%%%%%%%%%%%%%%%%%%%%%%%%%%%%%%%%
 The model membrane is expected to swell out at large $\alpha$ and crumple at small $\alpha$. Therefore, the shape of membrane, which is an extrinsic property of surfaces, is characterized by $X^2$. In order to see the correlation in the series of $X^2$, the autocorrelation coefficient
\begin{eqnarray}
A(X^2)= \frac{\sum_i X^2(\tau_{i}) X^2(\tau_{i+1})} 
   {  \left[\sum_i  X^2(\tau_i)\right]^2 },\\ \nonumber
 \tau_{i+1} = \tau_i + n \times 500, \quad n=1,2,\cdots  
\end{eqnarray}
is plotted in Fig. \ref{fig-1}(a), where $X^2(\tau_i)$ is sampled at every $n \times 500 \; (n=1,2,\cdots) $ MCS.  We easily see in Fig. \ref{fig-1}(a) that $A(X^2)$ at $\alpha\!=\!1400$ and the one at $\alpha\!=\!1600$ are completely different from each other. This result implies that the convergence speed of our MC on $N\!=\!1500$ fluid surfaces at $\alpha\!=\!1400$ is about 10 times slower than that at $\alpha\!=\!1600$, and indicates the existence of some discontinuous phase transition. 

The rate of acceptance $r_X$ for $X$ and $r_T$ for bond flip are plotted in Fig. \ref{fig-1}(b). We see in Fig. \ref{fig-1}(b) that $r_X$ is almost constant and $r_T$ is discontinuous. In order to maintain $0.5\leq r_X\leq 0.6$, we adjust $\epsilon$ in Eq. (\ref{epsilon}) to a suitable value, because $r_X$ rapidly changes against $\alpha$ at the transition point. This is the reason why $r_X$ is not completely constant. The discontinuous behavior of $r_T$ also indicates that the model undergoes a discontinuous transition.

%%%%%%%%%%%%%%%%%%%%%%%%%%%%%%%%%%%%%%%%%%%%%%%%%%%%%%%%%%%
\begin{figure}[htb]
\resizebox{0.49\textwidth}{!}{%
  \includegraphics{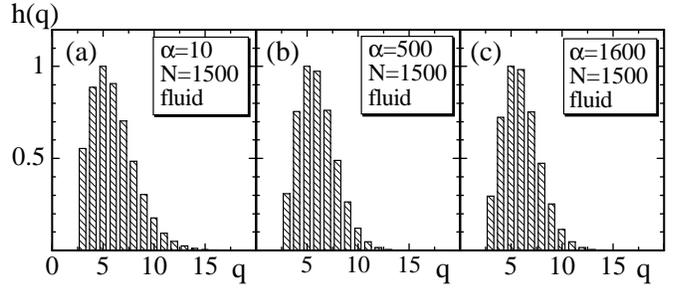}
}
 \caption{The normalized histograms $h(q)$ of the co-ordination number $q$ of the fluid surface of $N\!=\!1500$ at (a) $\alpha\!=\!10$ (crumpled phase), (b) $\alpha\!=\!500$ (branched polymer phase), and (c) $\alpha\!=\!1600$ (smooth phase).} 
\label{fig-2}
\end{figure}
%%%%%%%%%%%%%%%%%%%%%%%%%%%%%%%%%%%%%%%%%%%%%%%%%%%%%%%%%%%

In order to see the distribution of the co-ordination number $q$, the normalized histograms $h(q)$ are shown in Figs. \ref{fig-2}(a), \ref{fig-2}(b), and \ref{fig-2}(c). The histograms $h(q)$ in the figures were obtained in the final $5\!\times\!10^6$ MCS on the fluid surface of $N\!=\!1500$ at (a) $\alpha\!=\!10$, (b) $\alpha\!=\!500$, and (c) $\alpha\!=\!1600$. These three values of $\alpha$ correspond to three distinct phases; the crumpled, the branched polymer, and the smooth phases, which will be clarified later. We note that the co-ordination number $q$ does not always reach the above mentioned $q_i^{\rm max}$ the maximum co-ordination number during the final $5\!\times\!10^6$ MCS in each case. Thus, we find from Figs. \ref{fig-2}(a), \ref{fig-2}(b), and \ref{fig-2}(c) that the co-ordination number $q$ is dominated by $4\!\leq\!q\!\leq\!7$ at each $\alpha$ in the fluid model. No distinct difference can be seen in $h(q)$ obtained in the three phases, although it can bee seen that $h(q\!=\!3)$ in the crumpled phase at $\alpha\!=\!10$ is larger than those in the other two phases. It is remarkable that $h(q)$ in the smooth phase shown in Fig. \ref{fig-2}(c) is almost identical with $h(q)$ in the branched polymer phase shown in Fig. \ref{fig-2}(b), because we expect that the curvature energy $S_3$ is closely related with the co-ordination number $q$ and plays an essential role in the phase transition. The variation of $q$ is also very small even at the discontinuous transition point between the branched polymer and the smooth phases. 

Nevertheless, we can find in $S_3/N$ a discontinuous change, which reflects a discontinuous transition between the branched polymer and the smooth phases, in both models. However, the gap in $S_3/N$ is very small and comparable with the standard deviation of $S_3/N$. Hence, we can not conclude only from the gap that the model undergoes a discontinuous transition. 

%%%%%%%%%%%%%%%%%%%%%%%%%%%%%%%%%%%%%%%%%%%%%%%%%%%%%%%%%%%
\begin{figure}[htb]
\resizebox{0.49\textwidth}{!}{%
  \includegraphics{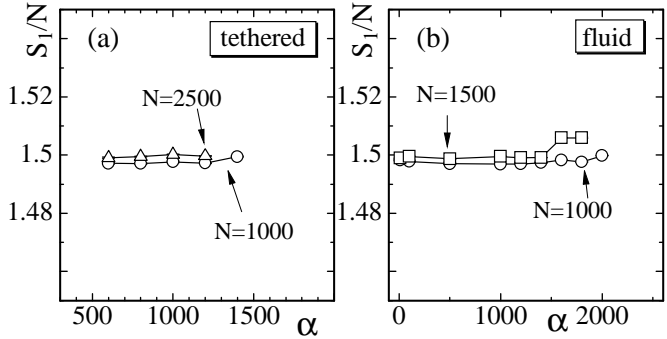}
}
 \caption{(a) $S_1/N$ obtained from $N\!=\!1500$ and $N\!=\!2500$ tethered surfaces, and (b) $S_1/N$ obtained from $N\!=\!1000$ and $N\!=\!1500$ fluid surfaces.} 
\label{fig-3}
\end{figure}
%%%%%%%%%%%%%%%%%%%%%%%%%%%%%%%%%%%%%%%%%%%%%%%%%%%%%%%%%%%
$S_1/N$ vs. $b$ are plotted in Figs. \ref{fig-3}(a) and  \ref{fig-3}(b). We see in Fig. \ref{fig-3}(a) that the expected relation $S_1/N\!=\!3/2$ is satisfied in the tethered model, and also in Fig. \ref{fig-3}(b) that $S_1/N\!=\!3/2$ in the fluid model.  The discontinuous change seen in $S_1/N$ of $N\!=\!1500$ in Fig. \ref{fig-3}(b) implies the existence of a discontinuous transition. Since the gap is almost equal to or less than $0.4\%$ of $S_1/N$, we consider that the relation $S_1/N\!=\!3/2$ is not influenced by the discontinuous transition. Thus, we understand that the surfaces in both sides of the transition point are in the equilibrium configurations prescribed by $Z$ of Eq. (\ref{Part-Func}).    

%%%%%%%%%%%%%%%%%%%%%%%%%%%%%%%%%%%%%%%%%%%%%%%%%%%%%%%%%%%
\begin{figure}[hbt]
\resizebox{0.49\textwidth}{!}{%
  \includegraphics{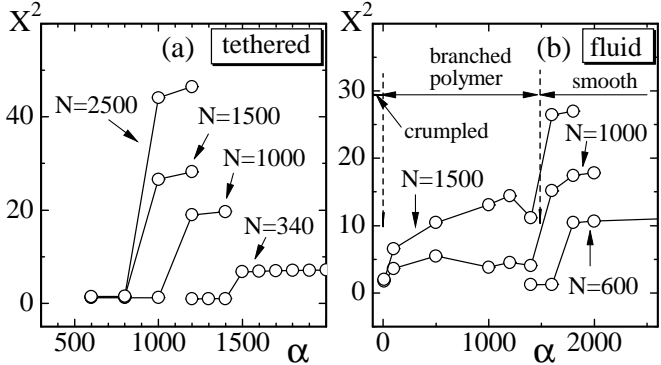}
}
\caption{$X^2$ vs $\alpha$ of (a) tethered surface model and of (b) fluid surface model. Dashed lines in (b) denote the phase boundaries on $N\!=\!1500$ fluid surfaces.}
\label{fig-4}
\end{figure}
%%%%%%%%%%%%%%%%%%%%%%%%%%%%%%%%%%%%%%%%%%%%%%%%%%%%%%%%%%%
The mean square size $X^2$ is plotted in Figs. \ref{fig-4}(a) and  \ref{fig-4}(b). In the tethered model, $X^2$ is obtained on surfaces of size up to $N\!=\!2500$. We see a gap in each $X^2$ in the tethered model. These discontinuous changes of $X^2$ imply that the phase transition is of first order. On the other hand, the dependence of $X^2$ on $\alpha$ in the fluid model appears to be continuous on surfaces of $N\!=\!1000$ and $N\!=\!1500$. The reason why $X^2$ takes intermediate values at $\alpha$ just below the transition point is that the surface belongs to a new phase, which emerges only in the fluid model. This new phase resembles the branched polymer phase \cite{KOIB-PRE-2003,Gom-Krol-2,Boal-Rao} in shape of the surfaces, which will be shown as a snapshot below. Thus, we call the phase as the branched polymer phase, although we can hardly confirm that the new phase is exactly identical with a branched polymer phase from the present numerical results. 

It should also be noted that the fluid surfaces become crumpled when $\alpha \!\to 0$, as seen in Fig. \ref{fig-4}(b). The dashed lines drawn vertically in Fig. \ref{fig-4}(b) denote the phase boundaries of the fluid surfaces of size $N\!=\!1500$. The leftmost data $X^2$ in Fig. \ref{fig-4}(b) were obtained at $\alpha\!=\!10$ on surfaces of $N\!=\!1000$ and $N\!=\!1500$, and their dependence on $N$ clearly different from those obtained at larger $\alpha$. Thus, we considered that the phase boundary between the crumpled and the branched polymer phases is located close at $\alpha\!=\!10$.     

%%%%%%%%%%%%%%%%%%%%%%%%%%%%%%%%%%%%%%%%%%%%%%%%%%%%%%%%%%%
\begin{figure}[htb]
  \resizebox{0.49\textwidth}{!}{%
  \includegraphics{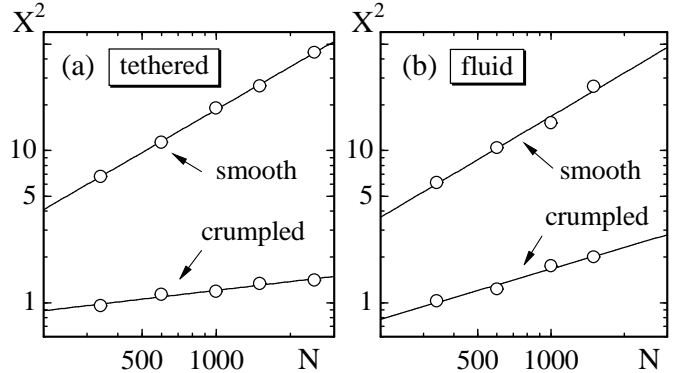}
}
\caption{$X^2$ vs $N$ on (a) tethered surfaces and on (b) fluid surfaces.  $X^2$ denoted by {\it smooth} in (a) and (b) were obtained in the smooth phase close at the transition point in both models. $X^2$ denoted by {\it crumpled} in (a) were obtained just below the transition point. $X^2$ of surfaces $N\!\leq \!600$ denoted by {\it crumpled} in (b) were obtained just below the transition point, and those of $N\!\geq\! 1000$ were obtained at $\alpha\!=\!10$. }
\label{fig-5}
\end{figure}
%%%%%%%%%%%%%%%%%%%%%%%%%%%%%%%%%%%%%%%%%%%%%%%%%%%%%%%%%%%
Figures  \ref{fig-5}(a) and  \ref{fig-5}(b) are log-log plots of $X^2$ against $N$.  $X^2$ denoted by {\it smooth} in the figures were obtained in the smooth phase close at the transition point in both models. $X^2$ denoted by {\it crumpled} in the tethered model were obtained just below the transition point. $X^2$ denoted by {\it crumpled} in the fluid model of $N\!\leq \!600$ were obtained just below the transition point, and those of $N\!\geq\! 1000$ were obtained at $\alpha\!=\!10$. The straight lines give the Hausdorff dimension $H$, which is defined by 
\begin{equation} 
X^2 \sim N^\frac{2}{H}.
\label{X2}
\end{equation} 
The surface swells and becomes almost smooth in the smooth phase both in the tethered model and in the fluid model; we already confirmed this from the data in Figs. \ref{fig-4}(a) and  \ref{fig-4}(b). $X^2$ has a value of the order of the radius squares when the membrane becomes a swollen sphere. Hence, we expect that $H\!\simeq \!2$ in the smooth phase. We have in fact 
\begin{eqnarray} 
\label{H}
&&H=2.11\pm 0.03 \quad ({\rm tethered}, {\rm smooth}),\nonumber \\
&&H=10.47\pm 1.23 \quad ({\rm tethered}, {\rm crumpled}),  \\
&&H=1.99\pm 0.07 \quad ({\rm fluid}, {\rm smooth}),\nonumber \\
&&H=4.15\pm 0.47\quad ({\rm fluid}, {\rm crumpled}).\nonumber
\end{eqnarray} 

The results in Eq. (\ref{H}) obtained in the tethered model indicate that the phase transition between the crumpled and the smooth phases is characterized also by a gap of $H$. The reason why $H$ is finite in the crumpled phase comes from the fact that $X^2$ are obtained close at the transition point, which is far distant from $\alpha\!=\!0$, where $H$ is expected to be infinite.     

In the fluid model, on the contrary, the dependence of $H$ on $\alpha$ in the branched polymer phase can not be extracted from the present numerical data. In order to see how $H$ depends on $\alpha$ in the branched polymer phase, large-scale simulations are necessary.  Nevertheless, it is possible that $H$ continuously changes against $\alpha$ in the fluid model, if the new phase is the branched polymer one. In fact, it is expected that $H\!\simeq\!2$ in the branched polymer phase \cite{KOIB-PRE-2003,Gom-Krol-2,Boal-Rao}. In addition, we see in Fig. \ref{fig-4}(b) that $X^2$ smoothly changes between the branched polymer phase and the crumpled phase.  

%%%%%%%%%%%%%%%%%%%%%%%%%%%%%%%%%%%%%%%%%%%%%%%%%%%%%%%%%%%
\begin{figure}[hbt]
\resizebox{0.49\textwidth}{!}{%
  \includegraphics{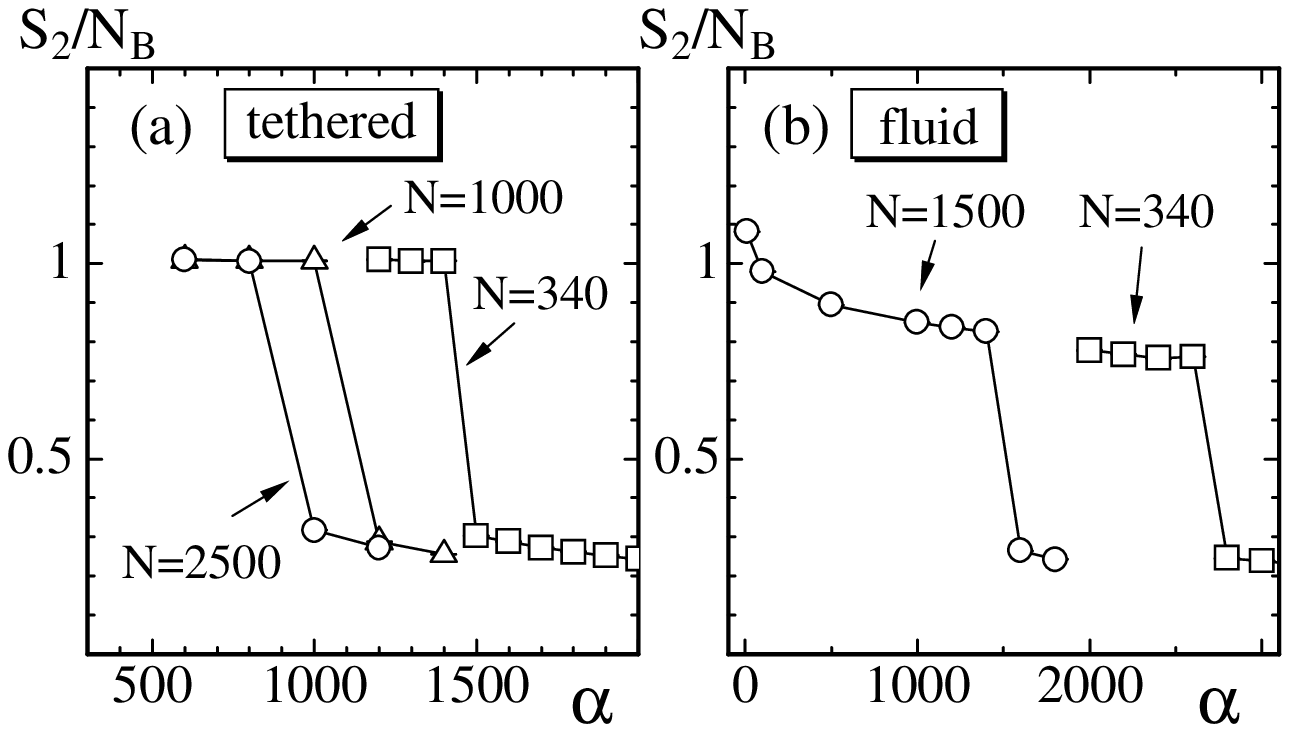}
}
\caption{$S_2/N_B$ vs $b$ obtained on (a) tethered surfaces and on (b) fluid surfaces. $S_2$ is defined by $S_2\!=\!\sum_i (1\!-\!\cos \theta_i)$.}
\label{fig-6}
\end{figure}
%%%%%%%%%%%%%%%%%%%%%%%%%%%%%%%%%%%%%%%%%%%%%%%%%%%%%%%%%%%
Figures  \ref{fig-6}(a) and  \ref{fig-6}(b) are the bending energy $S_2/N_B$ vs $\alpha$  respectively obtained on the tethered surfaces and on the fluid surfaces, where $S_2$ is defined by $S_2\!=\!\sum_i (1\!-\!\cos \theta_i)$,  $N_B$ the total number of bonds. $S_2$, which is an extrinsic variable and is not included in the Hamiltonian, represents how smooth the surface is. The gap seen in $S_2/N_B$ also indicates that the phase transition is of first order. Thus, we see a first-order transition between the smooth and the crumpled phase in the tethered model, and also see a first-order transition between the smooth phase and the branched polymer phase in the fluid model. 

We note that first-order nature of transitions is ordinarily confirmed by a discontinuous change (or a gap) of energy term included in Hamiltonian. A gap in some physical quantity that is not included in Hamiltonian can also confirm loosely a first-order nature of transitions. Thus, we consider that the gap in $X^2$ in Figs. \ref{fig-4}(a) and \ref{fig-4}(b), and the gap in $S_2$ in  Figs. \ref{fig-6}(a) and \ref{fig-6}(b) indicate the existence of  first order transitions.

%%%%%%%%%%%%%%%%%%%%%%%%%%%%%%%%%%%%%%%%%%%%%%%%%%%%%%%%%%%
\begin{figure}[h]
\centering{
  \resizebox{0.3\textwidth}{!}{%
\includegraphics{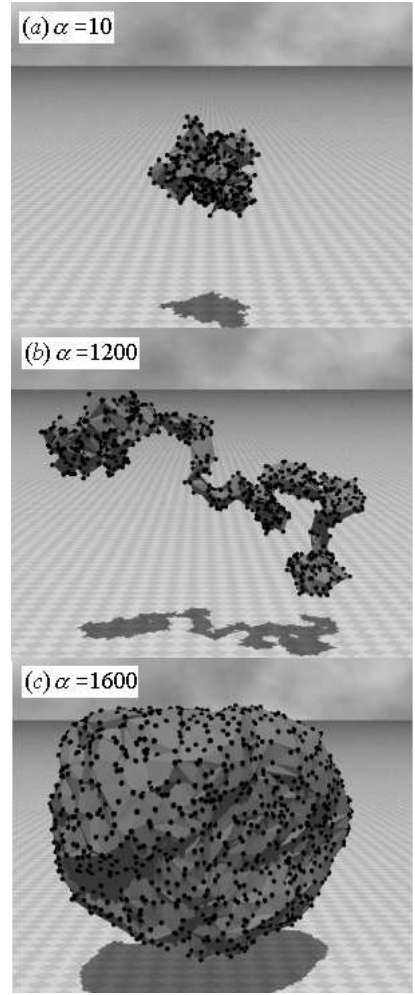}
}}
 \caption{Snapshots of fluid surface obtained in (a) the crumpled phase at $\alpha\!=\!10$, (b) the branched polymer phase  at $\alpha\!=\!1200$, and (c) the smooth phase  at $\alpha\!=\!1600$. Small spheres represent the vertices.  $N\!=\!1500$. }
\label{fig-7}
\end{figure}
%%%%%%%%%%%%%%%%%%%%%%%%%%%%%%%%%%%%%%%%%%%%%%%%%%%%%%%%%%%
Figures \ref{fig-7}(a), \ref{fig-7}(b) and \ref{fig-7}(c) are snapshots of surfaces obtained respectively at the smooth phase, at the branched polymer phase, and at the crumpled phase in the fluid surface model. The tethered surface model gives the same appearance of surfaces as the fluid surface model both in smooth and in crumpled phases, although the tethered model has no branched polymer phase.

%--------------------------------------------
\section{Summary and Conclusions}
%--------------------------------------------
We have studied extrinsic properties of surface models of membranes embedded in ${\bf R}^3$. The Hamiltonian contains the Gaussian energy term $S_1$ and the deficit angle term $S_3$; $S\!=\!S_1-\alpha S_3$. This $S_3$ was introduced by a straightforward extension of the co-ordination dependent term, which comes from the weight of the integration measure $dX$. Two kinds of models were investigated: one is a tethered model and the other is a fluid one. We found in the tethered model that there are two distinct phases; smooth and crumpled, and that these phases are connected by a discontinuous transition. It was also found in the fluid model that there is a new phase (branched polymer phase) between the smooth and the crumpled phases. A discontinuous transition separates the branched polymer phase from the smooth one in the fluid model. Although we called the new phase in the fluid model as the branched polymer phase, large scale simulations are necessary to confirm that the new phase is exactly identical with the branched polymer one.

%\acknowledgments
This work is supported in part by a Grant-in-Aid for Scientific Research, No. 15560160.

% BibTeX users please use
% \bibliographystyle{}
% \bibliography{}

\end{document}